\documentclass[useAMS,usenatbib]{mn2e}
\usepackage{graphicx,amsmath,multirow,amssymb}
\usepackage{subfig}
\usepackage{natbib}
\newcommand{\comment}[1]{}

\def\simgt{\lower.5ex\hbox{$\; \buildrel > \over \sim \;$}}
\def\simlt{\lower.5ex\hbox{$\; \buildrel < \over \sim \;$}}

\title[Galactic DC and OC PNe]{Galactic planetary nebulae with precise nebular
abundances as a tool to understand the evolution of asymptotic giant branch
stars}

\author[Garc\'{\i}a-Hern\'andez et al.]{D. A.~Garc\'{\i}a-Hern\'andez$^{1,2}$, P. Ventura$^3$, G. Delgado-Inglada$^4$, F. Dell'Agli$^{3}$, 
\newauthor
M. Di Criscienzo$^3$, A. Yag\"ue$^{1,2,3}$\\
$^1$Instituto de Astrof\'{\i}sica de Canarias, E-38205 La Laguna, Tenerife, Spain \\
$^{2}$Departamento de Astrof\'{\i}sica, Universidad de La Laguna (ULL), E-38206 La Laguna, Tenerife, Spain\\
$^{3}$INAF - Osservatorio Astronomico di Roma, Via Frascati 33, 00040, Monte Porzio Catone (RM), Italy \\
$^4$Instituto de Astronom\'{\i}a, Universidad Nacional Aut\'onoma de M\'exico, Apdo. Postal 70264,04510, M\'exico D. F., M\'exico 
}
\begin{document}

\date{Accepted, Received; in original form }

\pagerange{\pageref{firstpage}--\pageref{lastpage}} \pubyear{2012}

\maketitle

\label{firstpage}

\begin{abstract}
We present nucleosynthesis predictions (HeCNOCl) from  asymptotic giant branch
(AGB) models, with diffusive overshooting from all the convective borders, in
the metallicity range $Z_{\odot}/4 < Z < 2Z_{\odot}$. They are compared to
recent precise nebular abundances in a sample of Galactic planetary nebulae
(PNe) that is divided among double-dust chemistry (DC) and oxygen-dust chemistry
(OC) according to the infrared dust features. Unlike the similar subsample of
Galactic carbon-dust chemistry PNe recently analysed by us, here the individual
abundance errors, the higher metallicity spread, and the uncertain dust
types/subtypes in some PNe do not allow a clear determination of the AGB
progenitor masses (and formation epochs) for both PNe samples; the comparison is
thus more focussed on a object-by-object basis. The lowest metallicity OC PNe
evolve from low-mass ($\sim$1 M$_{\odot}$) O-rich AGBs, while the higher
metallicity ones (all with uncertain dust classifications) display a chemical
pattern similar to the DC PNe. In agreement with recent literature, the DC PNe
mostly descend from high-mass (M $\geq$3.5 M$_{\odot}$) solar/supersolar
metallicity AGBs that experience hot bottom burning (HBB), but other formation
channels in low-mass AGBs like extra mixing, stellar rotation, binary
interaction, or He pre-enrichment cannot be disregarded until more accurate C/O
ratios would be obtained. Two objects among the DC PNe show the imprint of
advanced CNO processing and deep second dredge-up, suggesting progenitors masses
close to the limit to evolve as core collapse supernovae (above $6~M_{\odot}$).
Their actual C/O ratio, if confirmed, indicate contamination from the third
dredge-up, rejecting the hypothesis that the chemical composition of such
high-metallicity massive AGBs is modified exclusively by HBB. 
\end{abstract}


\section{Introduction}
All the stars of mass in the range $1M~_{\odot} < M < 8~M_{\odot}$, after the
core He burning phase, evolve through the asymptotic giant branch (AGB) just
before the planetary nebula (PN) and white dwarf phases. AGB stars are mainly
supported by shell H burning but $3\alpha$ nucleosynthesis is
periodically activated in a He-rich layer above the degenerate core. These
periodic episodes are commonly referred to as thermal pulses because the
ignition of this nuclear channel occurs under condition of thermal instability
\citep{schw65, schw67}. The thermally-pulsing (TP) phase at the end of the AGB
is particularly important because nucleosynthesis primarily occurs in this
evolutionary stage. During the TP-AGB phase, the so-called third dredge-up (TDU)
brings the products of H and He burning to the stellar surface, increasing the
C/O ratio over unity and forming C-rich AGB stars. In the more massive (say, M
$>$ 3$-$4 M$_{\odot}$ at solar metallicity) AGBs, the star's surface may be also
enriched in N (and $^{13}$C) at the expenses of $^{12}$C as a consequence of hot
bottom burning \citep[HBB, e.g.][]{Mazzitelli1999}, keeping the C/O ratio below
unity. 

AGB stars play a significant role in several astrophysical contexts: i) they are
used to infer the masses of galaxies at high redshifts \citep{maraston06}
because of their large IR luminosities; ii) they provide an important
contribution to the light elements enrichment of the interstellar medium (ISM)
in our Milky Way Galaxy and other galaxies \citep{romano10} due to their high
gas pollution capability via stellar winds; iii) AGBs, because of the efficiency
of the dust formation process in their winds, play a crucial role in the
formation and evolution of galaxies \citep{santini14}, proving essential for the
understanding of the dust content in high-redshift quasars \citep{valiante11};
iv) the more massive HBB AGB stars \citep[e.g.][]{garcia06,garcia07,garcia09}
are currently believed to have provided the gas required to form
second-generation stars in globular clusters \citep{ventura01, dercole08}. The
above arguments represent only some examples, showing why the AGB evolutionary
phase, despite lasting only a tiny fraction of the whole life of a star, is at
the center of the astrophysical debate. Addressing the afore mentioned topics
requires full knowledge of the AGB evolution properties, and a detailed
description of the gas and dust yields provided by these objects.

AGB modelling has been significantly improved in the last few years; the last
generation of AGB models also include a description of the dust formation
process \citep{fg06, paperI, paperII, paperIII, paperIV, nanni13a, nanni13b,
nanni14}. This opens the possibility of determining the dust produced by AGBs,
in terms of the chemical composition of the dust particles formed, the dust mass
budget and the grain size distribution. The reliability of the results presented
so far, somewhat quantified through the differences among the model results of
the different groups involved in this research, is not satisfactory though. This
is primarily due to the poor knowledge of convection and mass loss, still
modelled via semi-empirical descriptions and that deeply affect the results
obtained \citep{vd05a, vd05b, doherty14a, doherty14b}. Because we are still far
from a self-consistent and physically sound treatment  of both mechanisms,
still based on first principles, the best way to make a significant step forward
into this direction is the comparison between the theoretical predictions and
the astronomical observations. 

The chemistry of PNe is the result of the several nucleosynthesis processes that
modify the stellar chemical composition during the previous AGB phase. Thus, PNe -
which can be observed at large distances and the nebular gas abundances can be
derived -  turn out to be particularly useful objects in order to test the AGB
theoretical models \citep{marigo03, marigo11, letizia09}. Some species such as C and
N \citep[to a more limited extend O; e.g.][]{gloria,garcia16} are greatly altered during
the earlier AGB phase, while another ones like Cl remain almost constant to values
typical of the original ISM. For these reasons we started a project dedicated to the
comparison between  the observations of PNe and the surface chemical composition of
AGB models in the latest evolutionary phases, with the ultimate goal of putting
additonal constraints on the description of the AGB phase.

In the first two papers of this series \citep{ventura15, ventura16} we focused
on the PNe sample of the Large and Small Magellanic Clouds (LMC and SMC); these
works offered an interesting opportunity to complete, from a different
perspective, the analysis focused on the interpretation of the {\it Spitzer}
sample of AGBs in the same galaxies \citep{flavia14, flavia15a, flavia15b}. The
analysis by \citet{ventura15, ventura16}, based on the observed chemical
composition, particularly of the CN abundances, allowed a characterisation of
the PNe observed in terms of formation epoch and progenitor's metallicity and
mass; this investigation provided interesting information regarding the
efficiency of the processes able to alter the surface chemistry of AGBs,
particularly at the low metallicities typical of the stars in the Magellanic
Clouds. 

In this paper we make a step forward, extending our analysis to higher
metallicities and comparing our new model predictions at solar and supersolar
metallicities with the sample of Galactic PNe recently studied by
\citet{gloria}. The \citet{gloria} work is based on high-quality optical spectra
in conjunction with the best available ionization correction factors (ICFs),
which allowed an accurate determination of the nebular abundances of He, C, N,
O, and Cl. It is to be noted here that: i) their spectra are deep enough to
detect weak lines (such as the O recombination lines) and have a higher
resolution ($<$4 \AA; adequate to avoid blends) than most works in the
literature (usually $>$7 \AA); ii) they used the most recent ICFs
\citep{gloria14}, which seem to work better (for most elements) than the
commonly adopted \citet{Kingsburgh94} ones; iii) apart from the usual
uncertainties from the line fluxes, they also considered the uncertainties
associated with the ICFs; this gives larger uncertainties  \citep[as compared to
the literature; e.g.][]{henry10,stan10,garcia14} but their total abundances are
more realistic; and iv) their sample PNe have available space-based mid-IR
spectra and can be classified depending on the dust features (e.g., C-rich or
O-rich), which provide an independent proxy for the nature of the PNe
progenitors. Indeed, we have recently compared the model predictions presented
here\footnote{\citet{garcia16} only gave a brief overview of the AGB ATON model
predictions presented here, leaving a more detailed presentation/discussion of
these models (especially the new solar/supersolar ones) to the present paper.}
with the \citet{gloria} sub-sample of low-metallicity Galactic PNe with C-rich
dust, obtaining a nice agreement between the models and the observational
results of O self enrichment in this type of PNe \citep{garcia16}\footnote{The
only nucleosynthesis models available at that time, predicting the O production
in some low-mass stars, were those by \citet{pignatari13} (only very recently
being accepted for publication).}. Thus, the goals of the present paper are the
detailed presentation of the new AGB model predictions for the HeCNO elements at
solar/supersolar metallicity as well as their comparison with the \citet{gloria}
samples of higher metallicity Galactic PNe with double- (both C- and O-rich) and
O-chemistry dust in their {\it Spitzer Space Telescope} and/or {\it Infrared
Space Observatory} mid-IR spectra. We describe the numerical and physical input
of the AGB models as well as the evolution of the surface chemistry (focussed on
the CNO elements) during the AGB in Sections 2 and 3, respectively. Section 4
presents the comparison and discussion of the PN nebular abundances with the
final chemical composition from the AGB evolutionary models, while our main
conclusions are given in Section 5.

\section{Numerical and physical input}
\label{input}
The AGB models used in the present analysis have been computed by means of the ATON
code for stellar evolution \citep{mazzitelli89}. A detailed description of the
numerical structure of the code is given in \citet{ventura98}, whereas the most
recent updates can be found in \citet{ventura09}.

The interested reader can find in the exhaustive reviews by \citet{herwig05}
and  \citet{karakas14} a detailed discussion on the main features of the AGB
evolution. The works by \citet{vd05a, vd05b} and \citet{doherty14a, doherty14b}
present a clear analysis of the uncertainties affecting the description of this
evolutionary phase, in particular how the  treatment of convection (both in terms of
the convective borders and the efficiency of the  convection model used), mass loss
and low-temperature molecular opacities reflect into the results obtained. 

The models used here were calculated with the use of the following  physical
ingredients:

\begin{table}
\begin{center}
\caption{Chemical and evolution properties of AGB models} 
\begin{tabular}{c|c|c|c|c|c|}
\hline
Z & Y & $[\alpha /Fe]$ & $M_C$ & $M_{HBB}$ & $M_{up}$  \\ 
\hline
$4\times 10^{-3}$  & 0.25   &  $+$0.2   &  1.1       &  3.5  &  6.0  \\ 
$8\times 10^{-3}$  & 0.26   &  $+$0.2   &  1.2       &  3.5  &  6.0  \\ 
0.018             & 0.28   &   0.0     &  1.4       &  3.5  &  5.5  \\ 
0.04              & 0.30   &   0.0     &  $\dots$   &  4.0  &  4.0  \\ 
\hline
\end{tabular}
\end{center}
\label{tabmod}
\end{table}

a) {\it Convection.} The convective instability was modelled according to the
full spectrum of turbulence (FST, hereafter) model, developed by \citet{cm91}.
In regions unstable to convection, mixing of chemical and nuclear burning are
coupled by means of a diffusion-like equation,  according to
\citet{cloutmann76}. Overshoot of convective eddies into radiatively stable 
regions is modelled via an exponential decay of velocities from the border of
the convective zones, fixed via the Schwartzschild criterion; the e-folding
distance of the decay is given by $\zeta H_p$\footnote{$\zeta H_p$ is the
e-folding distance of the exponential decay of convective velocities within
regions radiatively stable. If v$_{0}$ and HP$_{0}$ are the velocity and
pressure scale height at the formal border of convection, the velocity is
assumed to decay within the radiative zone as v=v$_{0}$$\times$exp[-r/($\zeta$HP$_{0}$)] where r is the distance from the
convective border.}. Following the calibration of the luminosity function of 
carbon stars in the LMC done in \citet{paperIV}, we use $\zeta=0.002$  to 
mimic overshoot from the base of the convective envelope and from the borders
of the  convective shell forming at the ignition of each TP.

b) {\it Mass loss.} To model mass loss, we used the recipe by \citet{blocker95} for
M-stars; this  description is based on hydrodynamical models of the envelope of
O-rich AGBs \citep{bowen88},  accounting for pulsation and the effects of
radiation pressure on dust grains. Concerning carbon stars, we used the results
from the Berlin group \citep{wachter02, wachter08}, which also consider dust
formation and the consequent effects of radiation pressure on carbonaceous dust
particles.

c) {\it Molecular opacities.} The molecular opacities in the low-temperature
regime  (below $10^4$ K) were calculated by means of the AESOPUS tool, developed
by \citet{marigo09}. With this approach the opacities are suitably constructed
to follow the alteration of the  chemical composition of the envelope,
accounting for changes in the individual abundances of C, N and O. This is
crucial for the description of the C-rich phase, because the increase in the
molecular opacities, occurring when the C/O ratio approaches (and overcomes)
unity, favours a considerable expansion of the surface layers of the star, with
the consequent enhancement of the rate at which mass loss occurs \citep{vm10}.

\begin{figure*}
\begin{minipage}{0.48\textwidth}
\resizebox{1.\hsize}{!}{\includegraphics{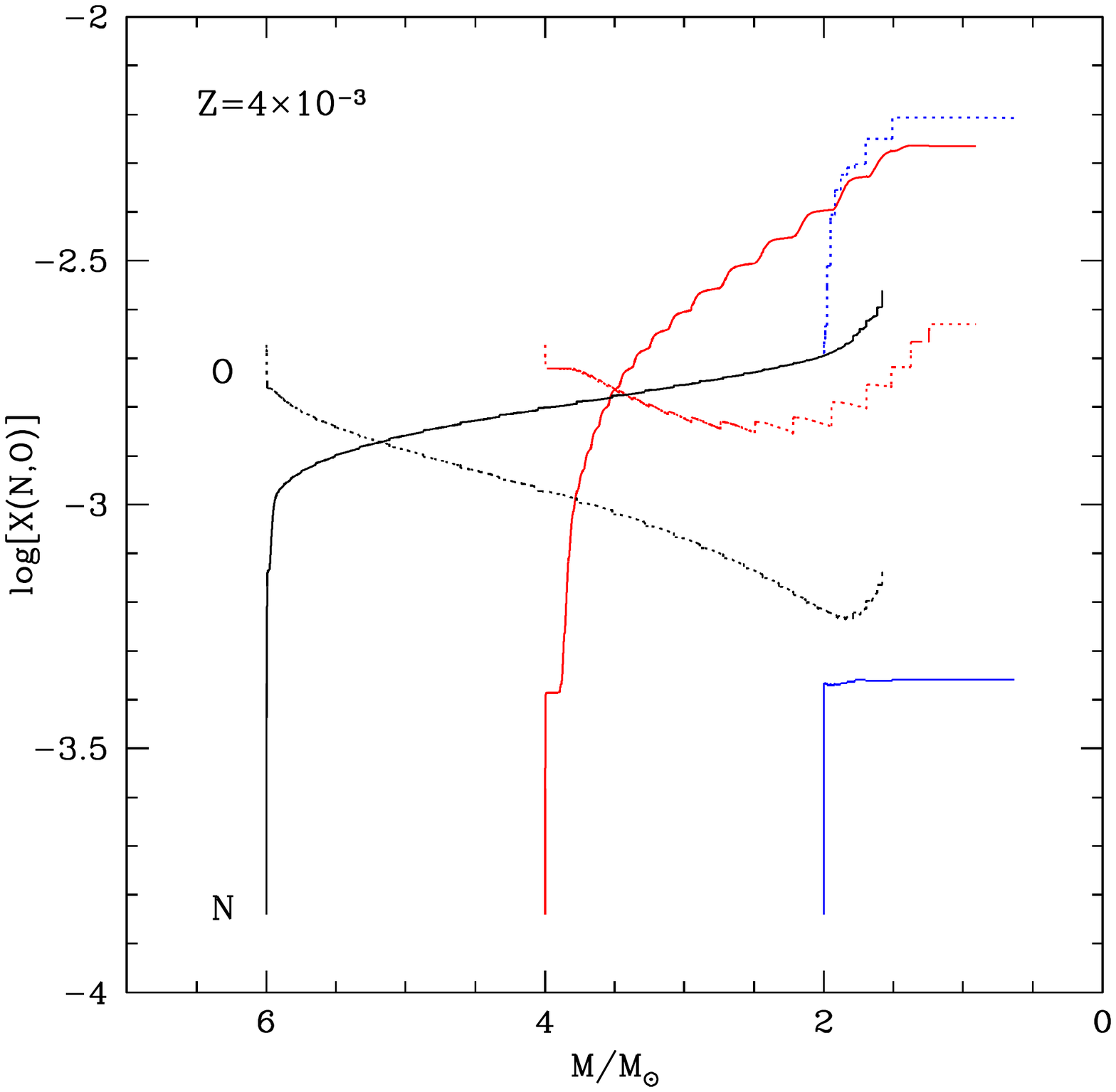}}
\end{minipage}
\begin{minipage}{0.48\textwidth}
\resizebox{1.\hsize}{!}{\includegraphics{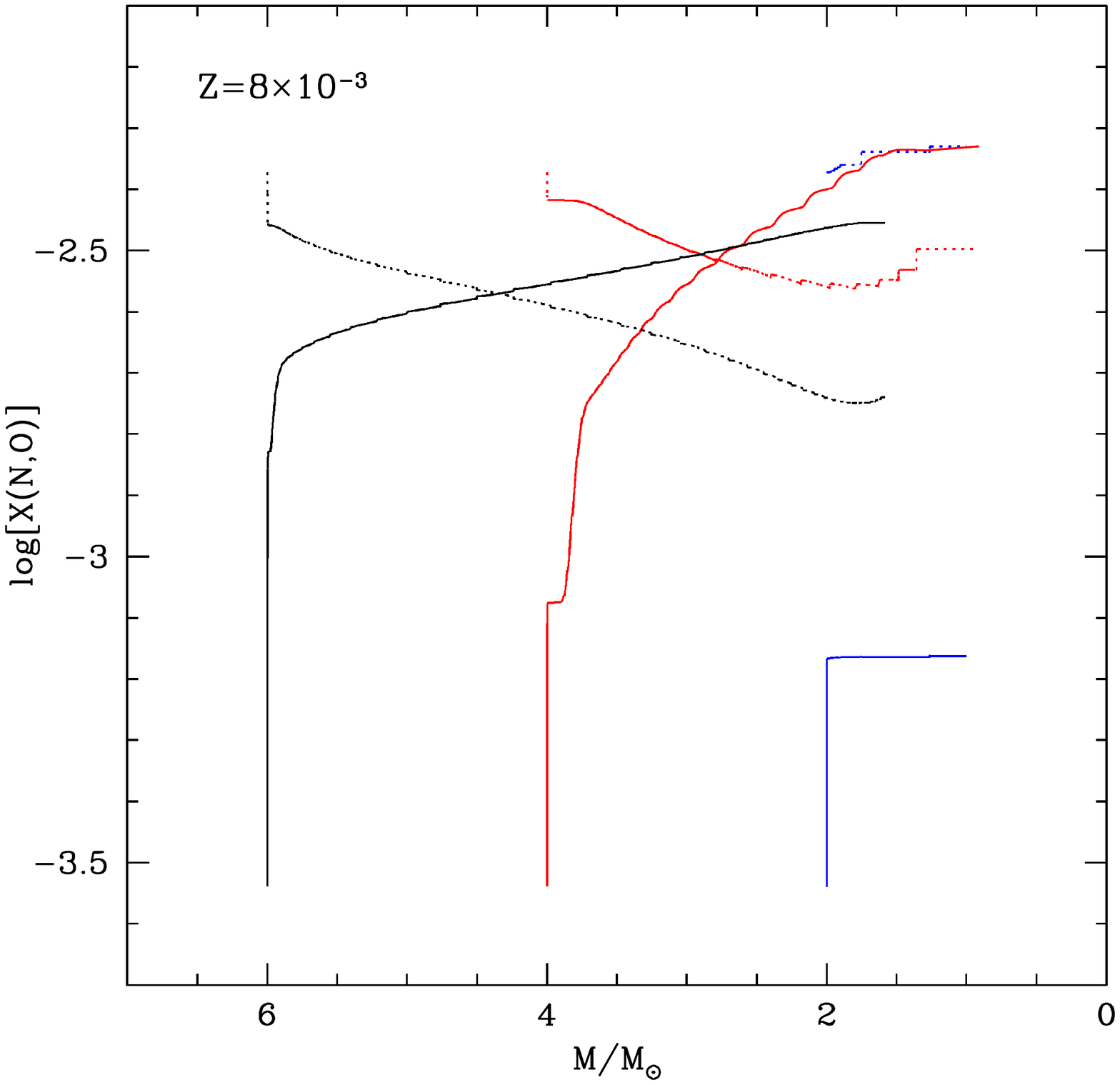}}
\end{minipage}
\vskip-70pt
\begin{minipage}{0.48\textwidth}
\resizebox{1.\hsize}{!}{\includegraphics{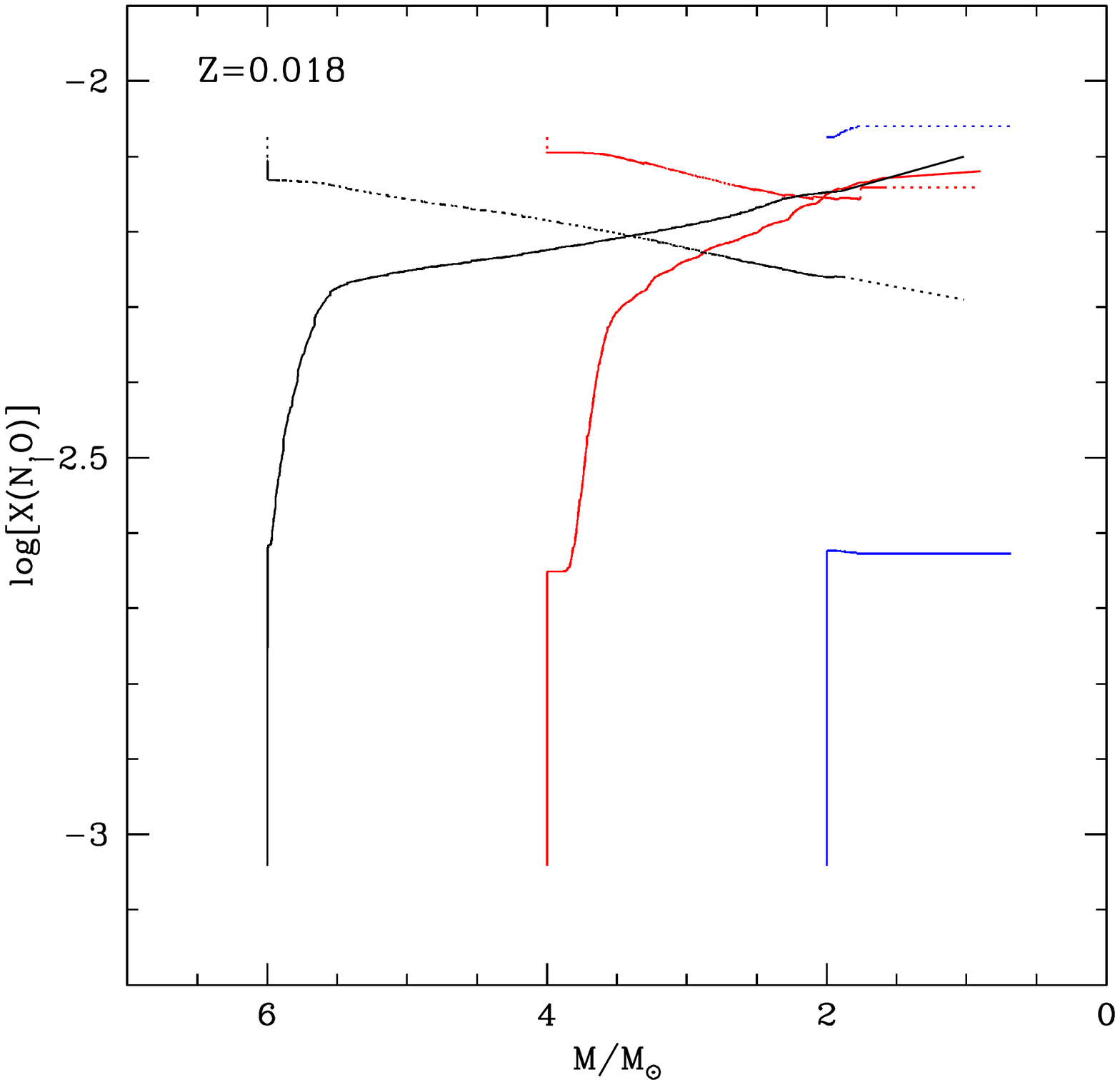}}
\end{minipage}
\begin{minipage}{0.48\textwidth}
\resizebox{1.\hsize}{!}{\includegraphics{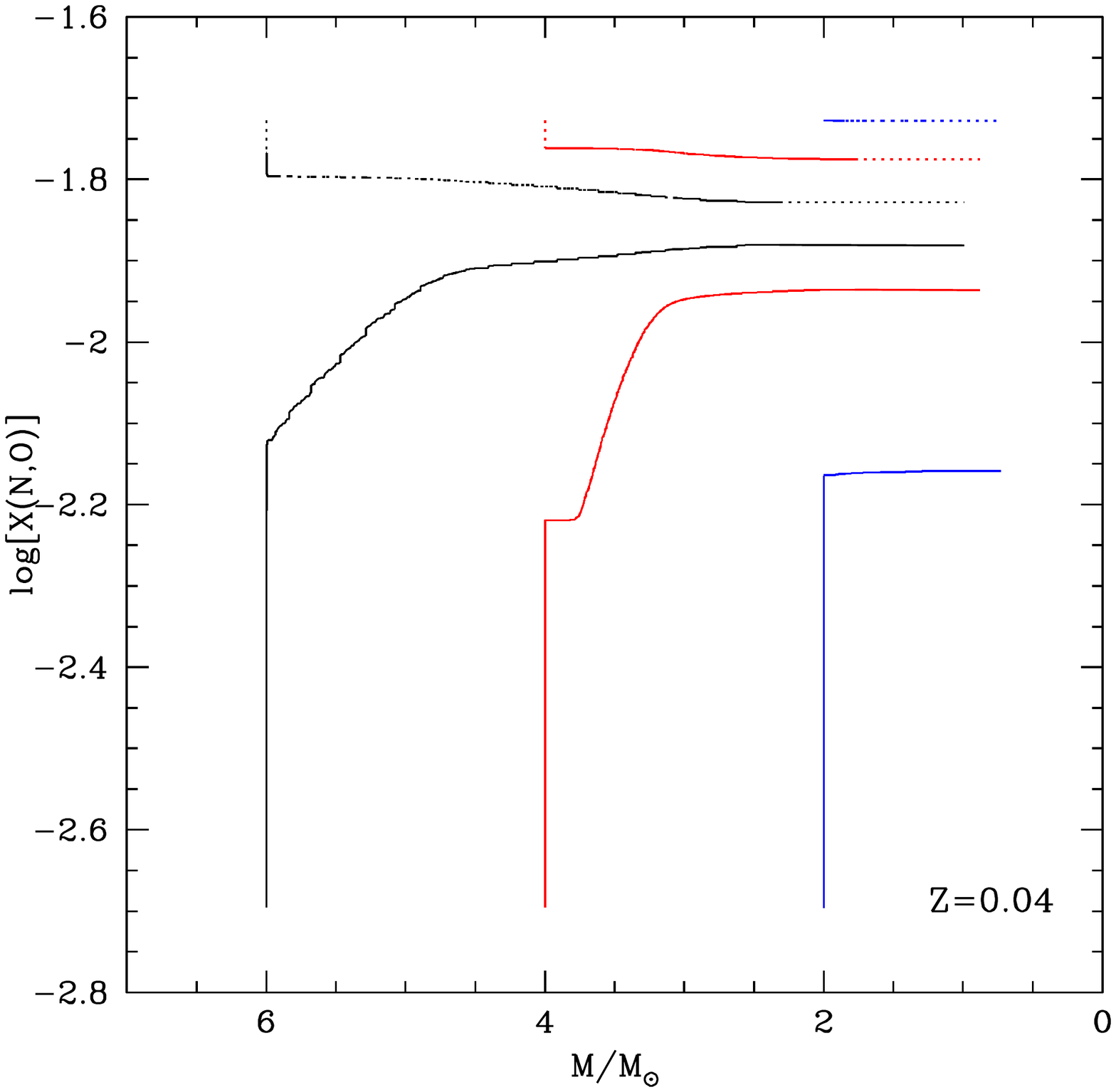}}
\end{minipage}
\vskip-40pt
\caption{The variation during the AGB phase of the surface mass fraction of N
(solid tracks) and O (dotted) in models of initial mass 2 (blue), 4 (red) and
$6~M_{\odot}$ (black) and metallicity $Z=4\times 10^{-3}$ (left, upper panel),
$8\times 10^{-3}$ (right, upper panel), 0.018 (left, lower panel) and 0.04
(right, lower panel). The vertical scale is logarithmic, to allow a better
coverage of the whole range of N and O abundances involved. On the abscissa we
report the current mass of the star.}
\label{fchem}
\end{figure*}

In this work we use models of metallicity $Z=4\times 10^{-3}$, $8\times
10^{-3}$, 0.018 and 0.04. The range of the initial mass encompass all the models
evolving through the AGB, i.e. $1~M_{\odot} \leq M \leq 8~M_{\odot}$.  The
mixture of the $Z=0.018$ and $Z=0.04$ models is solar-scaled, whereas for the
two lower metallicity ones we use an $\alpha-$enhancement $[\alpha/Fe]=+0.2$;
the relative fractions of the various species are taken from \citet{gs98}. A
summary of the initial chemical composition of the four sets of models is
reported in Table 1.

The $Z=4\times 10^{-3}$ models used here are presented and discussed in
\citet{ventura14b}, whereas the evolutionary sequences of metallicity $Z=8\times
10^{-3}$  are extensively illustrated in \citet{ventura13} (for initial mass above
$3~M_{\odot}$) and \citet{paperIV} (low-mass models of initial mass below
$3~M_{\odot}$). The solar and supersolar models have been calculated appositely
for the present analysis and a more extensive discussion of the corresponding
evolutionary sequences will be published in separate papers; e.g. the interested
reader may find deeper explanations for the evolutionary sequences at solar
metallicity in Ventura et al. (in preparation).

\section{The evolution of the surface chemistry during the AGB phase}
\subsection{A general overview}
\label{general}

The surface chemical composition of AGB stars is altered by TDU and HBB \citep[see
e.g.][]{karakas14}. The former consists in  the inwards penetration of the stellar
mantle, occurring at the end of each TP: the surface convection reaches layers
touched by He-nucleosynthesis. Mixing of nuclearly processed matter with the
external regions  favours the increase in the surface content of  C (mainly) and O
(in minor quantities). The HBB is activated at the base of the convective envelope,
when the temperature in those regions exceeds $\sim 30$ MK. The activation of this
process requires core masses above $\sim 0.8~M_{\odot}$. The ignition of HBB favours
CN nucleosynthesis, with production of N via proton capture by C nuclei. For
temperatures above $\sim 80$ MK the whole CNO cycling is activated, which further
favours N production at the expenses of C and O.

Before entering the discussion of how the efficiency of these mechanisms depends on
the mass and metallicity of the stars, we stress here that the results are highly
sensitive to the treatment of the convective borders \citep[TDU, e.g.][]{herwig00}
and to the efficiency of the convective model adopted \citep[HBB,][]{renzini81,
blocker91}. The discussion below is based on the physical ingredients used to
calculate the evolutionary sequences given in Section \ref{input}.

\subsection{Changes in the chemistry of AGBs: the role of mass and metallicity}
\label{change}

The modification of the surface chemical composition during the AGB phase is
determined by the relative contributions of TDU and HBB. The efficiency of the two
mechanisms depends on the core mass of the star; it is thus extremely sensitive to
the initial mass of the precursors. We may distinguish four cases:

\begin{enumerate}

\item{Stars with initial mass above $M_{up} \sim 6~M_{\odot}$ experience strong HBB,
with an advanced proton-capture nucleosynthesis occurring at the bottom of the
convective envelope, at temperatures above $80$ MK. Owing to the choice of the
\citet{blocker95} description of mass loss, ignition of HBB, with the consequent
rise in the luminosity \citep{blocker91}, is accompanied by a considerable increase
in the rate at which mass is lost, which, in turn, makes the star to experience a
small number of TPs, thus limiting the effects of TDU. The chemistry of these stars
is entirely determined by HBB: the overall C+N+O is preserved, N is greatly
enhanced, while part of C and O are destroyed by proton fusion. Because these stars
are those experiencing the most penetrating second dredge-up
\citep[SDU,][]{ventura10}, their envelope is enriched in He. The threshold mass,
$M_{up}$, depends on the metallicity of the star, because lower Z models, for a
given initial mass, evolve on bigger cores; the values of $M_{up}$ are shown in
Table 1.}

\item{Stars with mass in the range $M_{HBB} < M < M_{up}$ evolve with core mass
sufficiently large (above $0.8~M_{\odot}$) to experience HBB. Unlike their
counterparts  of higher mass (see point i above) they also experience TDU. Indeed in
the very final AGB  phases, when HBB is shut down by the consumption of the external
mantle, TDU becomes the  only mechanism able to alter the surface chemical
composition. In low-metallicity models, these late TDU episodes may convert O-rich
objects into C stars.  The final chemistry of these stars will be affected by both
HBB and TDU. While N will be produced in all cases, C and O may be created or
destroyed, according to whether the dominant mechanism is, respectively, TDU or HBB.
Here we also expect some He enrichment, though in smaller quantities in 
comparison with the previous case. $M_{HBB}$ changes with the metallicity of the
star  (see Table 1), because HBB is started more easily in models of lower
metallicity.}

\item{Stars with mass below $M_{HBB}$ do not experience any HBB, because  their core
mass are not sufficiently massive for the temperature at the bottom of the surface
envelope to reach the minimum value ($\sim 30$ MK) required to ignite HBB. TDU is
the only mechanism active in changing the surface chemical composition, thus
provoking a considerable increase in the surface C and, at a smaller extent, in the
O content \citep[see e.g.][]{garcia16}. A minimum threshold mass, $M_C$, exists,
above which the stars reach the C-star stage; lower mass AGBs consume their
envelope before the $C/O>1$ condition is achieved. $M_C$ is lower the lower is Z
(see Table 1), because: a) TDU is more efficient in lower-Z models
\citep{boothroyd88a, boothroyd88b};  b) when the metallicity is low, the star
contains a lower amount of O, thus a smaller quantity of C is needed to reach the
C-star stage.}

\item{Models with initial mass below $M_C$ never reach the C-star stage. These
objects evolve as O-rich stars; their surface chemistry is altered solely by
the first dredge-up (FDU), while ascending the red giant branch.}

\end{enumerate}

\begin{figure}
\centering
\includegraphics[width=0.48\textwidth, trim = 30 15 40 0, clip =yes]{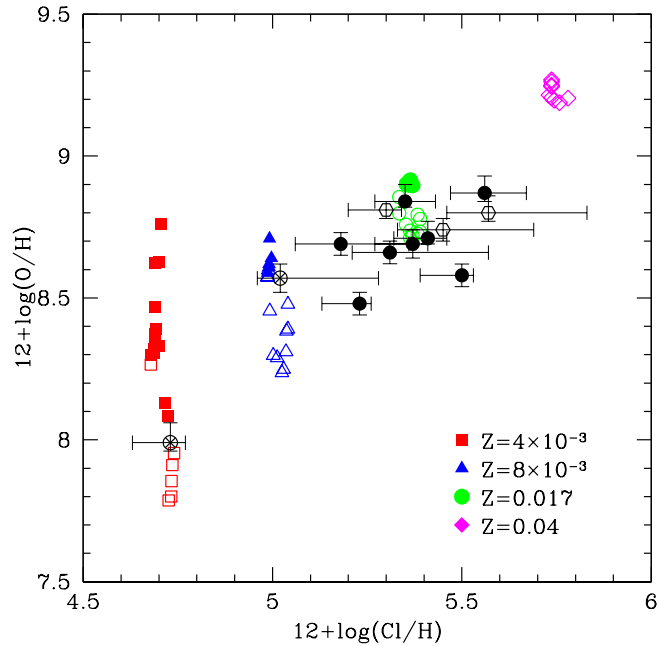}
\caption{The distribution of the Cl abundances versus the O ones in the
double-chemistry (DC) and oxygen-chemistry (OC) PNe by \citet{gloria} and of
the final abundances of the AGB models at several metallicities. The DC and
OC PNe are shown with solid and open-crossed circles, respectively, while the
three OC PNe (M 2-42, NGC 3132 and NGC 6543) with more uncertain dust
classifications (see text for more details) are shown with open hexagons.
Full(open) model points indicate carbon(oxygen)-rich chemistry.}
\label{fclo}
\end{figure}

\subsection{The evolution of the CNO elements}

Fig. \ref{fchem} shows the variation of the surface abundances of N and O in the
AGB models used here. For each Z we show the tracks of stars of initial mass  M
= 6, 4 and 2 $M_{\odot}$, taken as representative of the stellar groups (i),
(ii) and (iii), respectively, as introduced in Section
\ref{change}\footnote{Because the C-star stage is never reached in the Z=0.04
case, the $M=2~M_{\odot}$ model shown in the right, bottom panel of Fig.
\ref{fchem} is in fact representative of stars belonging to group (iv).}. We do
not show low-mass stars belonging to group (iv), because the corresponding
lines would show the effects of the FDU alone, causing a mere raise in the N
content. On the abscissa we report the current mass of the star: the tracks
start from the value of the initial mass, and move rightwards as mass is lost,
via stellar winds, from the envelope.

\begin{figure*}
\begin{minipage}{0.48\textwidth}
\resizebox{1.\hsize}{!}{\includegraphics{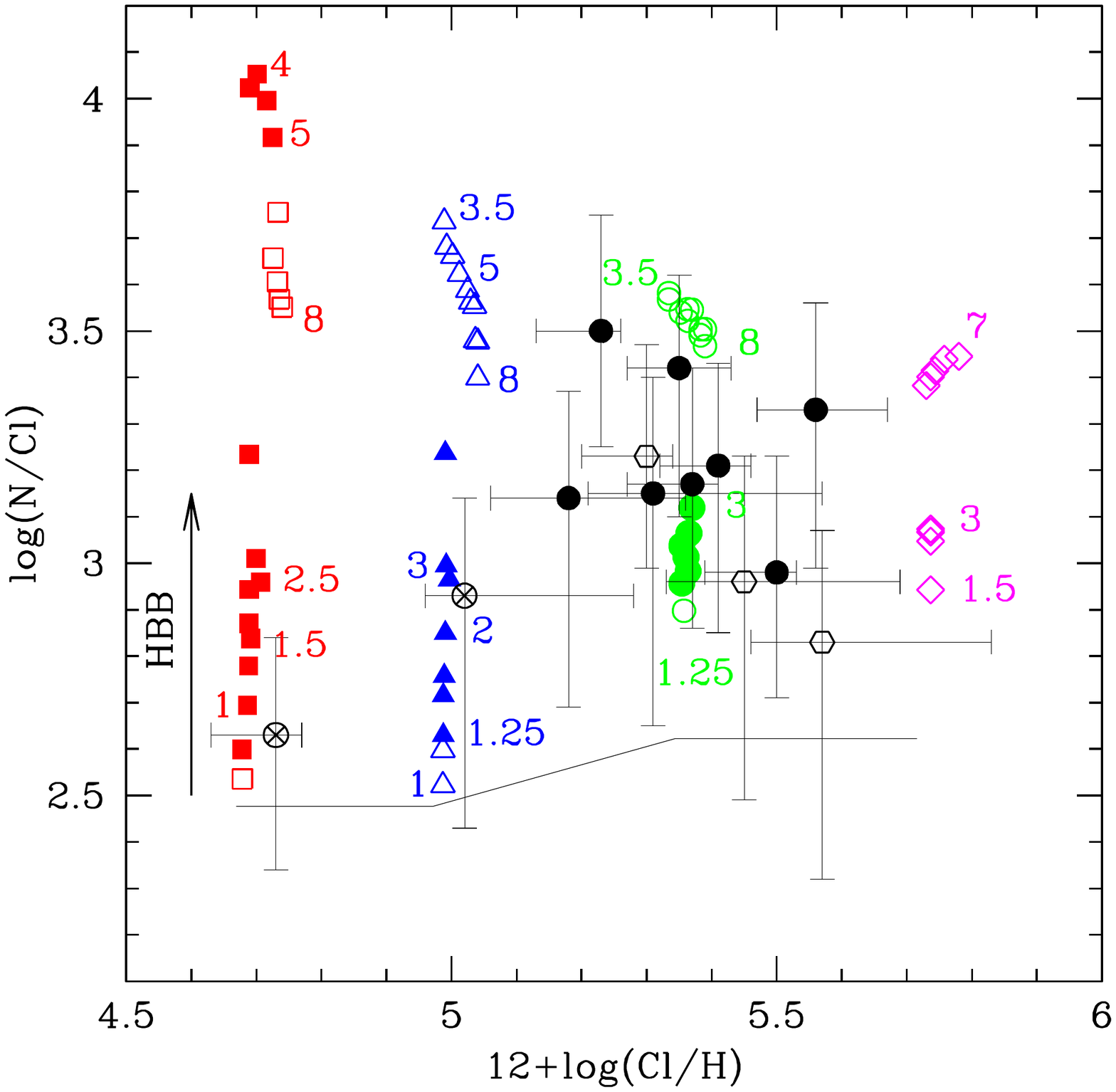}}
\end{minipage}
\begin{minipage}{0.48\textwidth}
\resizebox{1.\hsize}{!}{\includegraphics{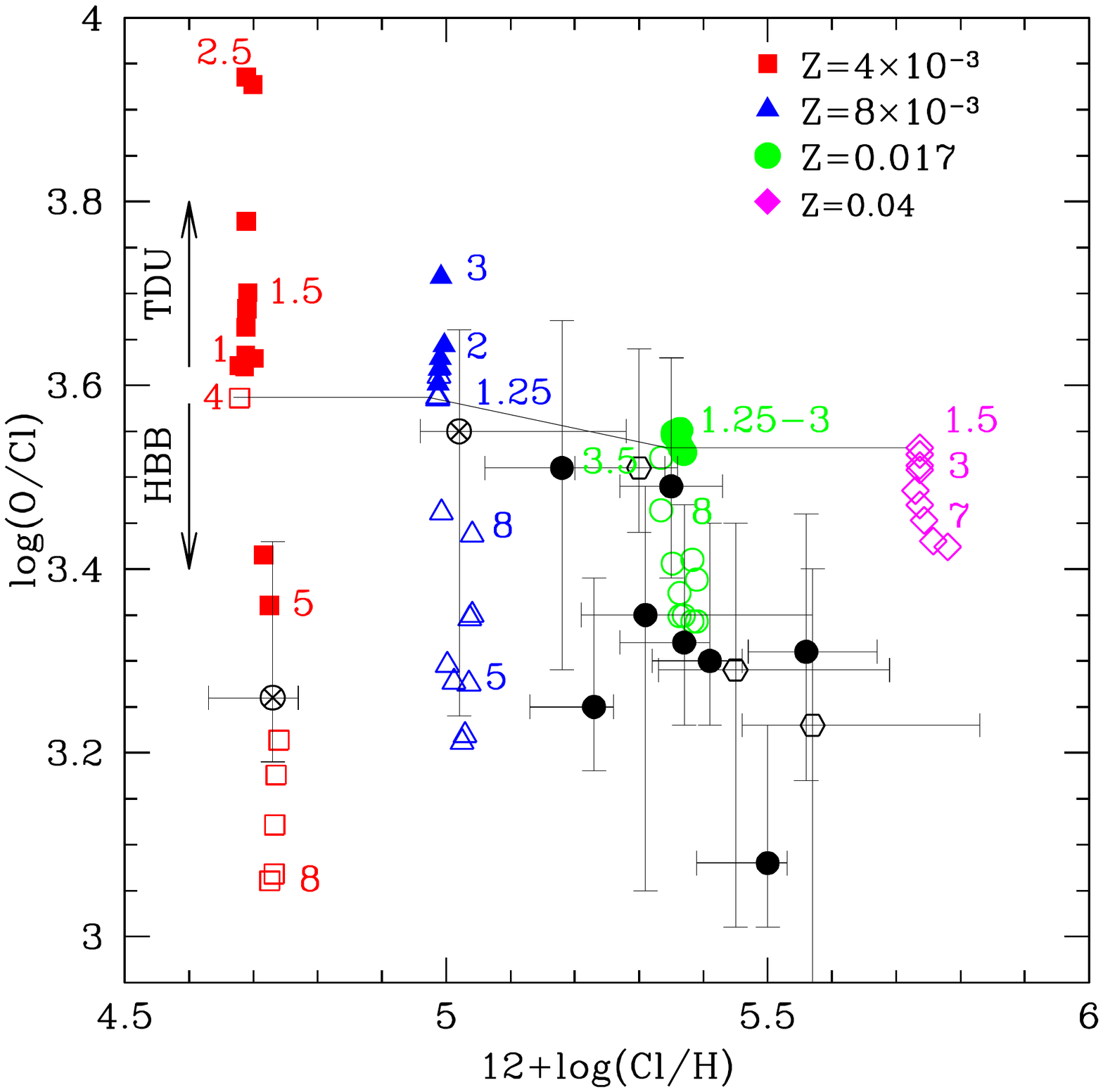}}
\end{minipage}
\vskip-70pt
\begin{minipage}{0.48\textwidth}
\resizebox{1.\hsize}{!}{\includegraphics{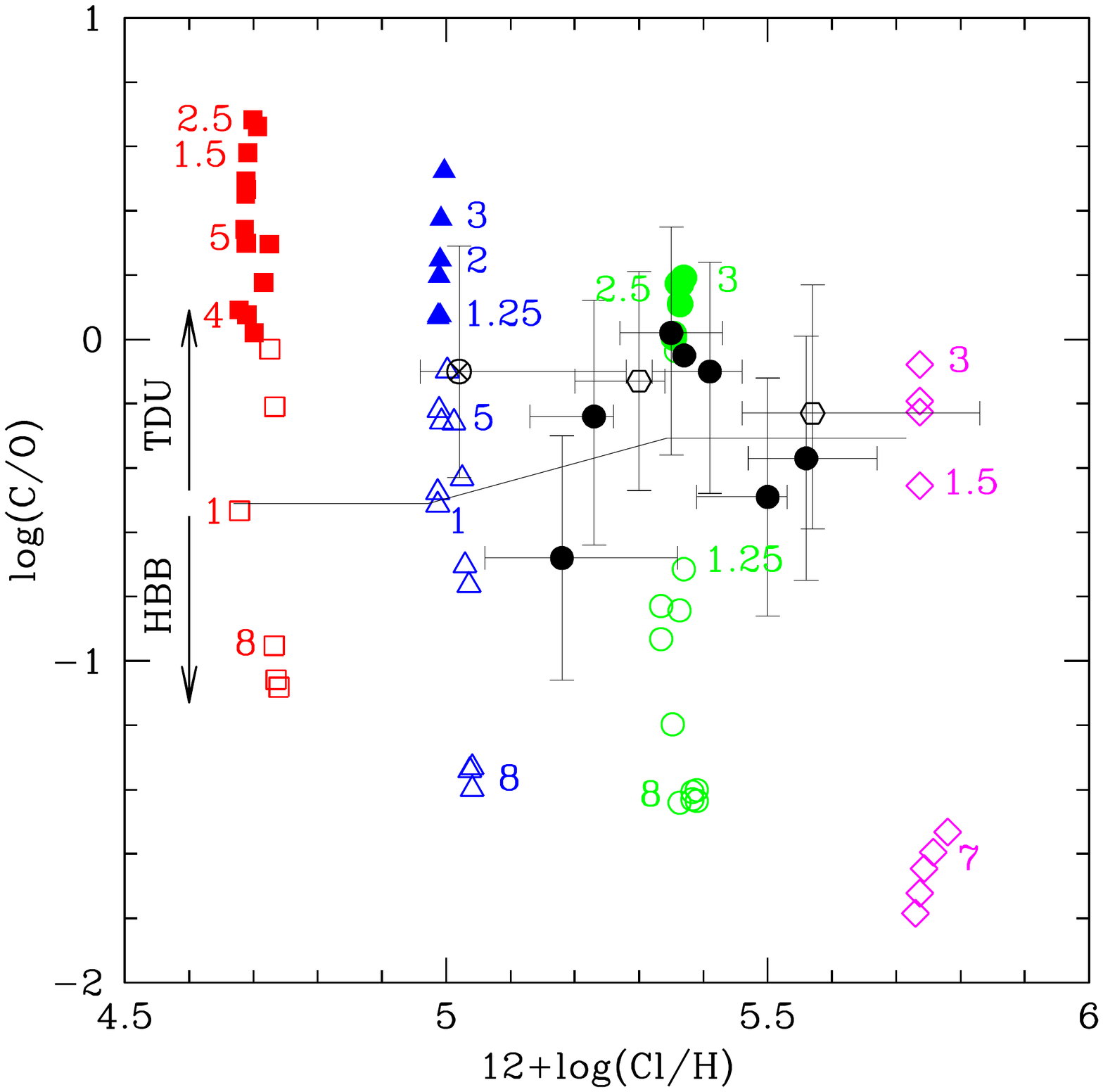}}
\end{minipage}
\begin{minipage}{0.48\textwidth}
\resizebox{1.\hsize}{!}{\includegraphics{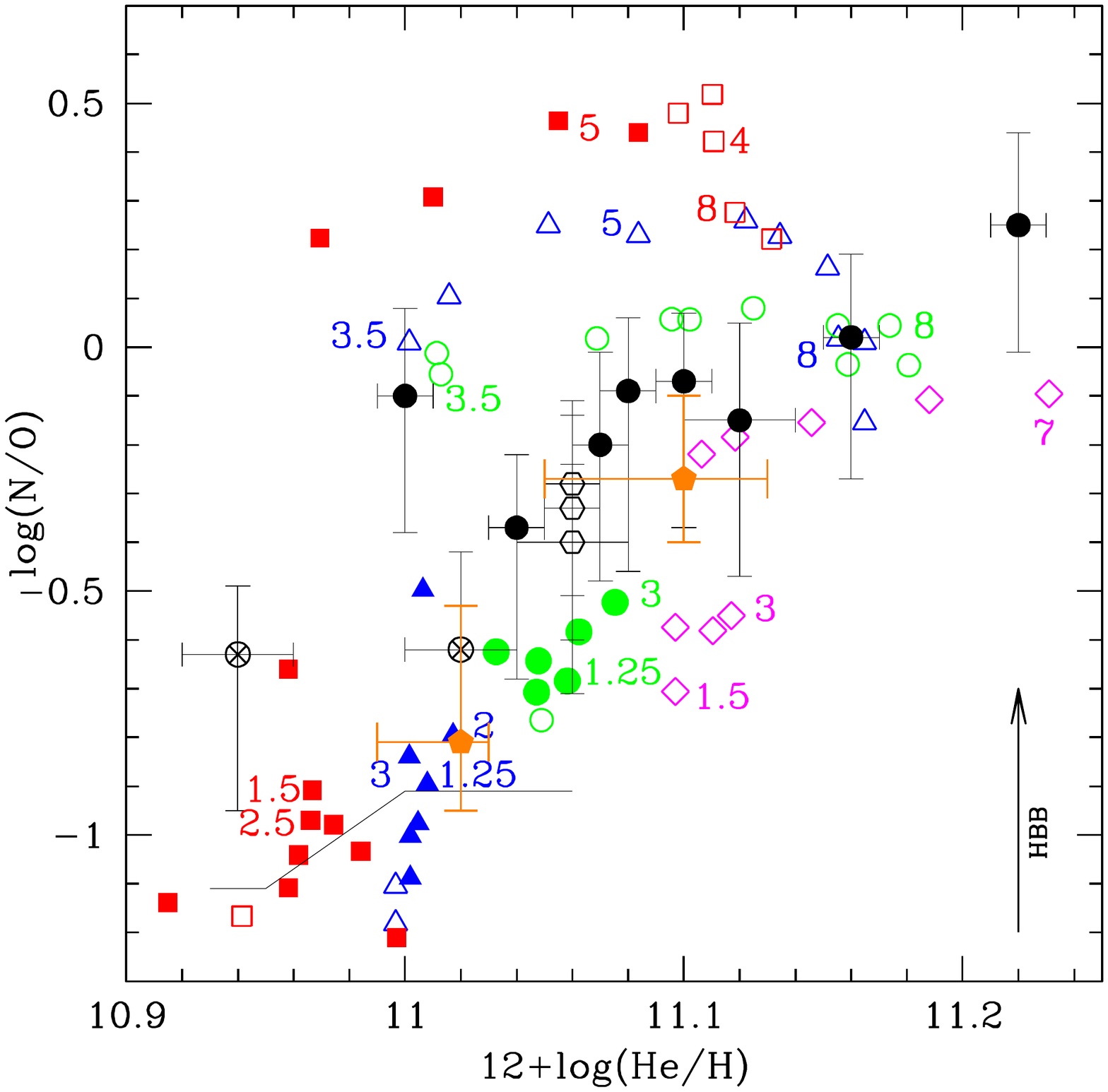}}
\end{minipage}
\vskip-40pt
\caption{The distribution of the chemistry of the double chemistry (DC) and
oxygen-chemistry (OC) PNe by \citet{gloria} and of the final mass fractions of
the AGB models of various metallicities in the Cl vs. N/Cl (top, left panel), Cl
vs. O/Cl (top, right), Cl vs. C/O  (bottom, left) and He vs. N/O (bottom, right)
planes. The DC and OC PNe are shown with solid and open-crossed circles,
respectively, while the three OC PNe (M 2-42, NGC 3132 and NGC 6543) with more
uncertain dust classifications (see text for more details) are shown with open
hexagons. Full(open) model points indicate carbon(oxygen)-rich chemistry.
The thin, solid lines indicate the assumed initial abundances in the models. The
arrows indicate the qualitative effect of HBB and of the SDU and TDU. The
orange pentagons (bottom-right panel) are the median abundances (and their
observed range as 25 and 75 percentiles) of Galactic OC (at
12+log(He/H)$\simeq$11.02) and DC (at 12+log(He/H)$\simeq$11.1) PNe measured by
\citet{garcia14}.} \label{fpne}
\end{figure*}

The chemistry of $6~M_{\odot}$ stars is entirely determined by HBB, with the drop of
the surface O and the synthesis of N. The surface C (not shown) is severely reduced
in these massive AGB models. Lower-Z models, experiencing a stronger HBB, undergo a
more significant variation in the surface chemical composition. The depletion in the
surface O during the whole AGB phase (see Fig. \ref{fchem}) amounts to  $\delta
\log(O) \sim -0.6$ for  $Z=4\times 10^{-3}$, $\delta \log(O) \sim -0.4$ for
$Z=8\times 10^{-3}$,  $\delta \log(O) \sim -0.25$ for $Z=0.018$ and $\delta \log(O)
\sim -0.1$ for $Z=0.04$. The corresponding increase in N is by a factor of $\sim 20$
in the $Z=4\times 10^{-3}$ model, $\sim 15$ in the $Z=8\times 10^{-3}$ case, $\sim
8$ for $Z=0.018$ and $\sim 5$ for $Z=0.04$

In $2~M_{\odot}$ models the change in the surface chemical composition occurs as a 
consequence of repeated TDU events. The rise of the surface C (by far in excess of
the O enhancement) may eventually turn an O-rich object into a C star, once the C/O
ratio exceeds unity. This occurs in all the  $2~M_{\odot}$ models shown in Fig.
\ref{fchem}, with the exception of the $Z=0.04$ case. For these stars, not
experiencing any HBB, the extent of the variation in the surface  chemical
composition is very sensitive to $Z$ (for the reasons given in Section
\ref{change})  and to the initial mass: the higher is the mass, the higher the
number of TDU episodes,  the larger the overall amount of C and O transported to the
star's surface. Therefore, the largest increase in the surface C is found in
models whose initial mass is close to $M_{HBB}$: for these stars, the C enhancement
factors are $50~(Z=4\times 10^{-3})$, $20~(Z=8\times 10^{-3})$, $5~(Z=0.018)$
and $2~(Z=0.04)$. Concerning O, the increase in the $Z=4\times 10^{-3}$ and
$Z=8\times 10^{-3}$ models of mass close to $M_{C}$ are, respectively, by a factor 3
and 2. No significant O increase is found in the higher-Z models. As for what
concerns N, the N content of low-mass AGBs increases by $\delta \log (N)
\sim 0.4$ dex (see Fig. \ref{fchem}), during the FDU, whereas no additional
variation is expected during the AGB phase. 

Stars of mass in the range $M_{HBB} < M < M_{up}$ (case (ii) in Section
\ref{change}) experience both TDU and HBB. This can be clearly seen in the
$4~M_{\odot}$  tracks in Fig. \ref{fchem} (particularly in the two upper panels,
corresponding to the lower metallicities), where we can distinguish three
phases: a) the initial AGB evolution, when the surface O decreases due to HBB
effects; b) an intermediate phase, during which O is destroyed by HBB in the
interpulse period, whereas it is transported to the surface by TDU; c) the final
AGB phase, when HBB is switched off, thus the surface O increases as a
consequence of the sole TDU effects. Nitrogen increases during the whole AGB
evolution of this class of models. Note that the N increase in the envelope is
larger in this case than in their higher mass counterparts because not only the
C initially present in the star is used to produce N via proton-capture, but
also because of the primary C dredged-up from the ashes of the He burning
shell.
 
\subsection{The final chemistry of AGBs}
\label{final}

While the temporal evolution of the surface chemistry of AGBs allows to
calculate the yields of the different elements, the interpretation of the PNe
chemistry requires the surface mass fractions of the various species at the end
of the AGB phase.

The final chemistry of the models used in the present analysis is shown in the
four panels of Fig.\ref{fpne}. Besides the CNO elements and He, we also show the
Cl abundance, because it is used as a good metallicity indicator by
\citet{gloria}. The latter element is not expected to experience any processing
during the AGB phase, remaining constant during the full stellar life. The
surface Cl is therefore the same as in the gas from which the star formed; thus
the theoretical predictions depend on the choice of the intial mixture. The
plots in Fig.\ref{fpne} were obtained based on the solar and $\alpha-$enhanced
mixture given by \citet{gs98}. We note, however, that the He abundance and
the abundance ratios used here (e.g., N/O, N/Cl, C/O) are consistent with
similar models calculated with the more recent \citet{asplund09} solar
abundances; this is also the case for the Monash AGB models (Karakas 2013, priv.
comm.). The massive AGB models eventually tend to the HBB equilibrium values,
which are primarily determined by the temperature at the base of the convective
zone, with scarce effect of the initial chemical composition. This holds in
particular for the CNO elements, which are fully involved in the nuclear
activity at the base of the envelope. For the less massive AGB stars, reaching
the C-star stage, what matters is the amount of C and O dredged-up to the
stellar surface. The latter is largely determined by the increase in the
mass-loss rate that occurs after becoming a C-star, while the initial chemistry
is completely forgotten.

The top panels of Fig.\ref{fpne} show the final abundances of N and O. The range of
N and O is more extended in models of lower metallicity because, as outlined in
Section \ref{change}, both HBB and TDU are more efficient in lower-Z environments.

The range of C/O values covered by the models is similar for the four
metallicities, while the range of the individual C and O abundances is extremely
sensitive to Z; the differences cancel when the C/O ratio is computed. The
largest C/O (over a factor of 10 higher than in the initial mixture) is found 
in the $Z=4\times 10^{-3}$ models, for the same arguments given in section
\ref{change}.

The distribution of the chemical composition in the N/O versus He plane
(right-bottom  panel of Fig.\ref{fpne}) is less straightforward. Generally
speaking, N/O is an indicator of the strength of HBB, while He is related to the
inwards penetration of the convective envelope during the SDU. The efficiency of
the latter mechanism, provoking the increase in the surface He, increases with the
mass of the star \citep{ventura10}. The distribution of the models in this plane is
dychotomic. Low mass AGB stars trace an approximately vertical sequence, with
constant He - no SDU is expected below $\sim 4~M_{\odot}$ and variable (though
always below $\sim 0.3$) N/O. Higher mass AGB models, experiencing HBB
are, however, spread on the right, upper region of the N/O-He plane.

\section{Understanding the Galactic PNe sample with precise nebular abundances}

The \citet{gloria} sub-sample (7 objects) of lower metallicity Galactic PNe with
C-dust features in their mid-IR spectra has been recently analysed by us
\citep{garcia16}. We note that, on average, the Galactic PNe with C-rich dust
features (in the form of aliphatic hydrocarbons), not analysed here, are of
lower metallicity than the double- and O-dust PNe. By comparing the observed
chemistry (HeCNOCl) with the ATON AGB models detailed here, \citet{garcia16}
concluded that they mostly descend from low-metallicity ($Z=8\times 10^{-3}$)
low-mass ($\sim$1$-$3 M$_{\odot}$) AGB stars that produce O, confirming
that O is not always a good metallicity indicator (especially in low-metallicity
C-rich dust PNe). In the present sample of OC and DC PNe (see below) the
situation is different \citep[significant O production/destruction is, in
principle, not expected;][see also Fig. \ref{fclo}]{gloria} and O is a more
reliable metallicity indicator. This is shown in Fig. \ref{fclo}, which displays
the Cl/H and O/H abundances; Fig. \ref{fclo}, however, is not the most useful to
distinguish the PNe progenitors (see Fig. \ref{fpne} for additonal abundance
ratios). 

The observed chemistry of the \citet{gloria} remaining sample (13 objects) of
Galactic PNe span a wide range of metallicities (by more than one order of
magnitude) as suggested by the Cl content (see below), which is taken as a
metallicity indicator\footnote{For consistency with the ATON AGB models, we
consider the solar Cl abundance of 12+log(Cl/H) = 5.50 \citep{gs98}.}. By
following the \citet{garcia14} dust type/subtype nomenclature, the available
{\it ISO} and/or {\it Spitzer} mid-IR spectra \citep[][see e.g. their Table
6]{gloria14}  are classified into two major dust types (oxygen chemistry or OC
and double chemistry or DC) and subtypes - amorphous ({\it am}) and crystalline
({\it cr}) - depending on the nature of the dust features; O-rich (OC) or both
C- and O-rich (DC). Eight out of the thirteen sources in the sample are
surrounded by both C-rich and O-rich dust, in the form of polycyclic aromatic
hydrocarbon (PAH) features and amorphous and/or crystalline silicates,
respectively. Five sources only display traces of amorphous ({\it am}) and/or
crystalline ({\it cr}) silicate features (O-rich dust) in their IR spectra.
Table 2 lists our Galactic PNe sample together with their IR dust types/subtypes
as well as the main abundance ratios used in this work. 

\begin{table*}
\centering
\begin{minipage}{180mm}
\caption{Main abundance ratios${^a}$ and IR dust types/subtypes in our Galactic PNe sample.} 
\begin{tabular}{l|c|c|c|c|c|c|c|l|l|}
\hline
Object & 12+log(Cl/H) & 12+log(O/H) & log(N/Cl) & log(O/Cl) & log(C/O) & log(N/O) & 12+log(He/H) & Dust type${^b}$ & Source${^c}$  \\ 
\hline
&  &  & & DC PNe &  &  &  &  \\ 
\hline
Cn 1-5   & 5.35$\pm0.08$          & 8.84$_{-0.02}^{+0.06}$ &  3.42$_{-0.32}^{+0.20}$ & 3.49$_{-0.10}^{+0.14}$ &  0.02$_{-0.38}^{+0.33}$  & -0.07$_{-0.26}^{+0.18}$ & 11.10$\pm0.01$ & DC$_{cr}$    & ISO/Spitzer \\ 
H 1-50   & 5.18$_{-0.12}^{+0.18}$ & 8.69$\pm0.04$          &  3.14$_{-0.39}^{+0.29}$ & 3.51$_{-0.16}^{+0.22}$ & -0.68$\pm0.38$           & -0.37$_{-0.31}^{+0.15}$ & 11.04$\pm0.01$ & DC$_{am+cr}$?& Spitzer    \\ 
M 1-42   & 5.23$_{-0.10}^{+0.03}$ & 8.48$\pm0.04$          &  3.50$_{-0.32}^{+0.18}$ & 3.25$_{-0.14}^{+0.07}$ & -0.24$_{-0.40}^{+0.36}$  &  0.25$_{-0.26}^{+0.19}$ & 11.22$\pm0.01$ & DC$_{cr}$    & ISO/Spitzer \\ 
M 2-27   & 5.56$_{-0.09}^{+0.11}$ & 8.87$_{-0.03}^{+0.06}$ &  3.33$_{-0.32}^{+0.25}$ & 3.31$_{-0.12}^{+0.17}$ & -0.37$\pm0.38$           &  0.02$_{-0.26}^{+0.20}$ & 11.16$\pm0.01$ & DC$_{cr}$    & Spitzer	 \\ 
M 2-31   & 5.31$_{-0.10}^{+0.26}$ & 8.66$\pm0.04$          &  3.15$_{-0.34}^{+0.41}$ & 3.35$_{-0.14}^{+0.30}$ &$\dots$                   & -0.20$_{-0.28}^{+0.19}$ & 11.07$\pm0.01$ & DC$_{cr}$    & Spitzer	 \\ 
MyCn 18  & 5.50$_{-0.11}^{+0.03}$ & 8.58$\pm0.04$          &  2.98$_{-0.35}^{+0.17}$ & 3.08$_{-0.15}^{+0.07}$ & -0.49$\pm0.37$           & -0.10$_{-0.28}^{+0.18}$ & 11.00$\pm0.01$ & DC$_{am+cr}$ & Spitzer	 \\ 
NGC 6439 & 5.37$_{-0.10}^{+0.04}$ & 8.69$\pm0.05$          &  3.17$_{-0.37}^{+0.19}$ & 3.32$_{-0.15}^{+0.09}$ & -0.05$_{-0.41}^{+0.37}$  & -0.15$_{-0.32}^{+0.20}$ & 11.12$\pm0.02$ & DC$_{cr}$    & Spitzer	 \\ 
NGC 7026 & 5.41$_{-0.09}^{+0.05}$ & 8.71$_{-0.02}^{+0.06}$ &  3.21$_{-0.40}^{+0.18}$ & 3.30$_{-0.11}^{+0.11}$ & -0.10$_{-0.38}^{+0.34}$  & -0.09$_{-0.33}^{+0.19}$ & 11.08$\pm0.01$ & DC$_{cr}$    & Spitzer	 \\ 
\hline
 &  &  & & OC PNe &  &   &  &  \\ 
\hline
DdDm 1   & 4.73$_{-0.10}^{+0.04}$ & 7.99$_{-0.03}^{+0.07}$ &  2.63$_{-0.35}^{+0.15}$ & 3.26$_{-0.13}^{+0.11}$ &$\dots$                   & -0.63$_{-0.28}^{+0.18}$ & 10.94$\pm0.02$ & OC$_{am}$    & Spitzer	  \\ 
M 2-42   & 5.45$_{-0.12}^{+0.24}$ & 8.74$\pm0.04$          &  2.96$_{-0.35}^{+0.39}$ & 3.29$_{-0.16}^{+0.28}$ &$\dots$                   & -0.33$_{-0.27}^{+0.19}$ & 11.06$\pm0.01$ & OC$_{cr}$?   & Spitzer	  \\ 
NGC 3132 & 5.30$_{-0.10}^{+0.04}$ & 8.81$\pm0.03$          &  3.23$_{-0.30}^{+0.18}$ & 3.51$_{-0.13}^{+0.07}$ & -0.13$\pm0.34$           & -0.28$_{-0.23}^{+0.17}$ & 11.06$\pm0.01$ & OC$_{cr}$?   & Spitzer  	\\ 
NGC 6210 & 5.02$_{-0.06}^{+0.26}$ & 8.57$\pm0.05$          &  2.93$_{-0.30}^{+0.41}$ & 3.55$_{-0.11}^{+0.31}$ & -0.10$_{-0.33}^{+0.39}$  & -0.62$_{-0.29}^{+0.20}$ & 11.02$\pm0.02$ & OC$_{cr}$    & Spitzer  	\\ 
NGC 6543 & 5.57$_{-0.11}^{+0.26}$ & 8.80$_{-0.03}^{+0.06}$ &  2.83$_{-0.36}^{+0.39}$ & 3.23$_{-0.14}^{+0.32}$ & -0.23$_{-0.36}^{+0.40}$  & -0.40$_{-0.28}^{+0.19}$ & 11.06$\pm0.02$ & OC$_{am+cr}$?& ISO	\\ 
\hline
\end{tabular}
${^a}$\ Abundance ratios and uncertainties from \citet{gloria}.\\
${^b}$\ IR dust type/subtype from \citet{gloria14} but following the nomenclature by \citet{garcia14}. Questions marks indicate a rather uncertain dust classification (see text for more details).\\
${^c}$\ The available {\it ISO} and/or {\it Spitzer} mid-IR spectra can be consulted in \citet{gloria14} (see also references in their Table 6). 
\end{minipage}
\label{tabmod}
\end{table*}

The DC-type PNe display Cl abundances from subsolar (12+log(Cl/H)$\sim$5.2) to
slightly supersolar (12+log(Cl/H)$\sim$5.6), while the OC-type ones span an even
larger Cl range, from very low-metallicity (12+log(Cl/H)$\sim$4.7; PN DdDm 1) to
supersolar values (12+log(Cl/H)$\sim$5.6; PN NGC 6543). Fig. \ref{fpne} shows
the chemical composition of our present Galactic PNe sample in comparison with
the AGB model predictions; for comparison we also show the median He and
N/O values of Galactic OC and DC PNe as measured by \citet{garcia14} from
low-resolution optical spectra and using the \citet{Kingsburgh94} ICFs\footnote{
\citet{garcia14} only could measured the Cl abundances in a few PNe and they
are very uncertain.}. Unlike the C-rich dust PNe \citep{garcia16}, here the
errors associated to the individual abundances (especially for the C/O ratio),
the higher spread in metallicity, and uncertainties in the dust type/subtype for
some objects (see below) do not allow a straight determination of the AGB
progenitor mass and the formation epoch of the DC and OC PNe samples. In any
case, some interesting conclusions may be extracted from the models versus
observations comparison, although it should be more foccused on a star-by-star
basis.

As we mentioned above, the O-dust chemistry PNe are generally more metal-rich
than their C-dust chemistry counterparts \citep{gloria,garcia16}. The only
exceptions are the halo PN DdDm 1 and NGC 6210\footnote{Based on the observed
morphology, \citet{soker16} very recently speculates that this PN could be the
result of triple-stellar evolution with a tight binary system.}, which show a
much smaller and a similar metallicity, respectively; DdDm 1 is also the only OC
PN showing amorphous silicates in emission (OC$_{am}$ in Table 2). Both objects
have a small N content, pointing against HBB contamination; this is also
confirmed by their low N/O ratios and He abundances. Their chemistries are
consistent with low-mass ($\sim$1 M$_{\odot}$), low-metallicity ($Z=4-8\times
10^{-3}$) progenitors, formed $\sim$ 5-6 Gyr ago\footnote{The formation epoch
estimates are according to ATON and are model dependent. In particular the ATON
evolutionary time-scales are usually shorter than other AGB models in the
literature such as those by Karakas (2010) \citep[see also Fig. 1
in][]{garcia13}.}, that did not reach the C-star stage (group (iv) in Section
\ref{change}). Our interpretation for PN DdDm 1, based on the ATON AGB models,
is fully consistent with the results from {\it Spitzer} mid-IR spectroscopy of
Galactic PNe \citep{letizia12} and their correlation with the chemical
abundances from low-resolution optical spectra \citep[basically He and the N/O
ratio;][see Fig. \ref{fpne}]{garcia14}, which show that Galactic OC$_{am}$ PNe display higher
Galactic latitudes than the OC$_{cr}$ ones and their He and N/O ratios are
consistent with their progenitors being the lowest metallicity and lowest mass
AGB stars in our Galaxy, respectively\footnote{A similar comparison for NGC 6210
is not possible because such low-metallicity OC$_{cr}$ objects are absent in the
\citet{garcia14} sample of PNe with {\it Spitzer} spectra.}. 

The other three O-rich dust PNe (M 2-42, NGC 3132 and NGC 6543) display higher
metallicities, between slightly subsolar and supersolar. We note, however, that
the dust classes for the OC PNe M 2-42 and NGC 3132 (both classified as
OC$_{cr}$, Table 2) and NGC 6543 (classified as OC$_{am+cr}$, Table 2) are more
uncertain. M 2-42 displays the lowest S/N IR spectrum in our sample and only
low-resolution (R$\sim$100) spectra are available, while NGC 3132 displays a
rather noisy IR spectrum, which looks different to the rest of IR spectra
(especially the slope of the underlying IR dust continuum emission between 10
and 38 $\mu$m). Both objects display a weak PAH-like feature at 11.3 $\mu$m that
seems to be not accompanied by the other PAH-like features at $\sim$6.2, 7.7,
and 8.6 $\mu$m \citep{gloria14}. On the other hand, NGC 6543 only has a rather
noisy {\it ISO} (much less sensitive than {\it Spitzer}) spectrum available
\citep{bernard05}, which hampers the detection of any dust feature in the
PAHs spectral region. This together with their nearly solar Cl abundances of
12+log(Cl/H)$=$5.30 (NGC 3132), 5.45 (M 2-42) and 5.57 (NGC 6543) indicate that
these PNe could be truly DC PNe where the PAH-like features may have escaped
detection by {\it Spitzer} and {\it ISO}; note that NGC 3132 is also suspected
to have a binary companion \citep[e.g.][]{sahu86}. \citet{garcia14} also found a
few similar examples of Galactic OC PNe with crystalline silicates (e.g.,
OC$_{cr}$) showing a chemical nebular gas pattern (e.g., He, Ar, and N/O)
identical to the one of DC PNe. Their possible link with DC PNe is also
suggested by our Fig. \ref{fpne} where the three objects (marked with open
diamonds) are indistinguishable from the rest of DC PNe in our sample.

The double-dust chemistry PNe are also more metal-rich than the C-dust chemistry
ones \citep{gloria,garcia16} and, on average, even more metal-rich
(12+log(Cl/H)$\geq$5.2) than the OC-type PNe (Fig. \ref{fpne}). Six out of the
eight DC sources only display crystalline silicates (DC$_{cr}$-type in Table 2).
The only exceptions are MyCn 18 and H 1-50, which are classified as DC$_{am+cr}$
PN because they also displays amorphous silicates in emission; although the
amorphous silicates emission in H 1-50 is very weak. It is to be noted here that
although Cn 1-5 and H 1-50 pertain to the DC class, they show {\it Spitzer}
spectra that are clearly different from the rest of DC PNe, which show the
typical DC spectrum with weak PAH-like bands and crystalline/amorphous
silicates. The Cn 1-5 IR spectrum display very strong PAH-like features and
unidentified 24 $\mu$m emission; curiously, this is the only DC object in our
sample with C/O$\geq$1 (but also consistent, within the errors, with C/O$<$1).
The Cn 1-5 C/O ratio is higher than one using both recombination lines (RLs) and
collisionally excited lines (CELs) \citep{gloria14}, suggesting a true C-rich
gas nebula. The H 1-50 IR spectrum, however, shows tentative PAH-like features
and the general shape of the IR spectrum is quite similar to other OC PNe such
as NGC 6210 or even DdDm 1. The PN H 1-50 may thus be a truly OC object. Indeed,
H 1-50 displays the lowest He and N/O ratios in the DC group, suggesting a less
massive progenitor than other objects of similar metallicity such as M 2-42.
Again, the individual abundances errors (especially for the key C/O ratio) and
the spread in metallicity difficult the determination of the AGB progenitor
masses of the DC PNe as a whole; exceptions to this limitation are PNe M 1-42
and M 2-27 (see below). The DC PNe almost cover all possible initial masses in
the Cl vs. N/Cl, Cl vs. O/Cl and Cl vs. C/O planes (Fig. \ref{fpne}). However,
the He vs. N/O plane (bottom, right panel in Fig. \ref{fpne}) suggests that,
according to the ATON AGB models, DC PNe are mostly the descendants of the
higher mass (M $\geq$3.5 M$_{\odot}$) AGB progenitors, experiencing HBB, at
solar/supersolar metallicity (formed $\sim$50-250 Myr ago). This is consistent
with \citet{gloria}, who compared with other AGB nucleosynthesis predictions in
the literature \citep[e.g., those from][]{karakas10}\footnote{\citet{gloria}
also used the model predictions by \citet{pignatari13}, although not accepted  at
that time.} and \citet{garcia14} where the median He and N/O abundances (from
low-resolution optical spectroscopy; see Fig. \ref{fpne}) of larger samples of Galactic disk and
bulge DC PNe were found to be consistent with the \citet{karakas10} predictions
for $\sim$5 M$_{\odot}$ solar metallicity HBB AGB stars\footnote{The minimum
mass to activate HBB is model dependent; e.g., at solar metallicity it is
$\sim$3.5 M$_{\odot}$ in our ATON AGB models, while it is $\sim$4.5 M$_{\odot}$
in the Karakas (2010) ones.}. However, as already pointed out by
\citet{garcia14} and \citet{gloria}, less massive ($<$3.5 M$_{\odot}$) non-HBB
stars could also produce the high He and N/O ratios observed in DC PNe via extra
mixing, stellar rotation, binary interaction, or even He pre-enrichment. 

Interestingly, MyCn 18 and M 1-42 (and M 2-27) seem to be examples of objects at
the lower and higher end, respectively, of the AGB progenitor masses covered by
DC PNe in the He vs. N/O plane. The PN MyCn 18 seems to be the descendant of a
solar metallicity $\sim$3.5 M$_{\odot}$ AGB progenitor, formed $\sim$250 Myr
ago. The chemical composition of PNe M 1-42 and M 2-27, however, clearly
reflects the effects of HBB, because they are enriched in N and have an O
content significantly smaller than the average value of other sample PNe at
similar metallicities. Furthermore, they are the two PNe with the highest He
abundances, suggesting that they descend from stars of mass $\sim 6-7M_{\odot}$
and metallicity slightly subsolar (M 1-42) and supersolar (M 2-27). The
evolutionary times of AGB stars of this mass and metallicity indicate a
relatively recent formation epoch, between 40 and 80 Myr ago. According to our
modelling, these stars should belong to group (i) in section \ref{change}, thus
their chemical composition should show-up the imprinting of HBB, with no
signatures of TDU. However, the recommended C/O ratios are at odds with this
hypothesis, as they are a factor $\sim 5$ higher than predicted (see left,
bottom panel of Fig.~\ref{fpne}), suggesting some contribution from TDU. This
possibility was advanced by \citet{garcia06,garcia07,garcia13}, to explain the
Rb overabundances observed in massive Galactic AGB stars, undergoing HBB.

Still in the context of massive AGBs, \citet{ventura15} outlined that some of
the PNe in the LMC can be interpreted as the progeny of $6-7~M_{\odot}$ stars,
whose surface chemistry was contaminated exclusively by HBB. Clearly the
comparison among the two samples is not straightforward, because the
metallicities of the PNe are different. However, confirmation of the C/O values
provided by \citet{gloria} would suggest larger effects from TDU in
solar/supersolar metallicity massive AGB stars. In the context of our modelling,
larger TDU effects would indicate that the mass-loss rate in our massive AGB
models of solar and supersolar metallicity is too high. Alternatively, the true
C/O ratios of M 1-42 and M 2-27 could be smaller than the \citet{gloria}
recommended values\footnote{\citet{gloria14} obtained a C/O ratio of 0.36 (from
CELs) and 0.56 (from RLs) and of only 0.32 (from RLs) for M 1-42 and M 2-27,
respectively, suggesting also that these abundance ratios are uncertain and
better spectra and calculations would be needed.}, thus rendering the agreement
with the models more satisfactory. 

In short, the comparison with the ATON models suggests that: i) the lowest
metallicity OC PNe should be the descendants of low-mass ($\sim$1 M$_{\odot}$)
stars that are not converted into C-rich stars, while the higher metallicity OC
ones (with uncertain dust classifications and DC-like chemical compositions)
could be truly DC PNe where the PAH-like features may have escaped detection in
the available mid-IR space-based spectroscopic observations; ii)
solar/supersolar metallicity DC PNe should be the descendants of the higher mass
(M $\geq$3.5 M$_{\odot}$) HBB AGB stars but alternative channels of formation
cannot be discarded with the present nebular abundances and their associated
errors. 

\section{Conclusions}

We use AGB models (with diffusive overshooting from all the convective borders)
of different metallicity (from $Z_{\odot}$/4 to 2Z$_{\odot}$) to interpret the
surface chemistry (He, C, N, O, and Cl) of a sample of Galactic PNe with
high-quality spectra, which together with the best ICFs available allowed
abundance determinations with unprecedented accuracy/reliability
\citep{gloria}. The PNe investigated are divided among double-dust chemistry
(DC) and oxygen-dust chemistry (OC) according to the dust features present in
their available space-based IR spectra. 

Unlike the \citet{gloria} subsample of carbon-dust chemistry (CC) PNe recently
analysed by us \citep{garcia16}, here the individual abundance errors (in
particular for the C/O ratio), the wider metallicity range, and the uncertain
dust types/subtypes in some objects do not permit a clear determination of the
AGB initial mass (and the formation epoch) for both PNe samples. The PNe
observations versus AGB models comparison is thus more focussed on a
star-by-star basis.

The two lowest metallicity OC PNe (DdDm 1 and NGC 6210) are interpreted as the
descendants of low-mass ($\sim$1 M$_{\odot}$) AGB stars that did not reach the
C-rich phase. The three higher metallicity PNe in this group (M 2-42, NGC 3132
and NGC 6543) have uncertain dust classifications and are chemically
indistinguishable from the rest of DC PNe in our sample, being interpreted as
truly DC PNe where the PAH-like features may have escaped detection by the {\it
Spitzer} and {\it ISO} satellites.

The DC PNe in our Galaxy may display different kind of IR spectra and,
sometimes, even C/O ratios over unity (e.g., Cn 1-5). However, we still lack
complete samples of Galactic DC PNe with precise C/O ratios. We find that the DC
PNe in our sample are best separated (in terms of progenitor AGB masses) in the
He vs. N/O plane, which otherwise suggests that they mostly descend from the
higher mass (M $\geq$3.5 M$_{\odot}$) HBB AGB stars at solar/supersolar
metallicity. This is consistent with recent works in the literature, and also we
cannot discard alternative formation channels in low-mass non-HBB stars such as
extra mixing, stellar rotation, binary interaction, or He pre-enrichment. More
precise C/O ratios turn out to be fundamental to learn about the stellar origin
of DC PNe.

The DC PN MyCn 18 seems to be at the lower end ($\sim$3.5 M$_{\odot}$) of the
AGB progenitor masses covered by the DC PNe in our sample. On the other hand,
two DC PNe (M 1-42 and M 2-27) are likely descendants of the more massive AGB
stars of close-to-solar chemistry, with mass $\sim 6-7~M_{\odot}$, formed
$40-80$ Myr ago. This is deduced on the basis of the large content of N and He,
and the low O. The recommended C/O ratio of these two objects ($C/O \sim 0.5$),
if confirmed, suggests a role by TDU in the contamination of the surface
chemistry during the previous AGB phase. This is at odds with our AGB model
predictions, giving a significantly smaller $C/O \sim 0.1$, with only a modest
contamination from TDU. Also, confirmation of this finding would indicate an
intrinsic difference among the evolution of massive AGBs, likely progenitors of
the DC PNe studied here, and their lower metallicity counterparts in the
Magellanic Clouds, which, as shown by \citet{ventura15, ventura16}, are
contaminated by HBB only. A more robust determination of the C/O ratio of these
two particular PNe is a promising opportunity to assess any possible
contribution of TDU in altering the surface chemical composition of massive AGB
stars in solar and supersolar environments.

\section*{Acknowledgments}
D.A.G.H. was funded by the Ram\'on y Cajal fellowship number RYC$-$2013$-$14182 and
he acknowledges support provided by the Spanish Ministry of Economy and
Competitiveness (MINECO) under grant AYA$-$2014$-$58082-P. P.V. was supported by
PRIN MIUR 2011 `The Chemical and Dynamical Evolution of the Milky Way and Local
Group Galaxies' (PI: F. Matteucci), prot. 2010LY5N2T. G.D.I. acknowledges support
from the Mexican CONACYT grant CB$-$2014$-$241732. F.D.A. acknowledges support from
the Observatory of Rome.

\end{document}